\documentclass[12pt]{article}

\usepackage{amsfonts}
\usepackage{amssymb}

\begin{document}

\title{Continuous ensembles and the $\chi$-capacity of infinite-dimensional
channels}
\author{A.S. Holevo, M.E.Shirokov\\Steklov Mathematical Institute, 119991 Moscow,
Russia}

\date{}
\maketitle

\section{Introduction}

This paper is devoted to systematic study of the classical
capacity (more precisely, a closely related quantity -- the
$\chi$-capacity) of infinite dimensional quantum channels,
following \cite{H-c-w-c}, \cite{H-Sh}, \cite{Sh}. While major
attention in quantum information theory up to now was paid to
finite dimensional systems, there is an important and interesting
class of  Gaussian channels, see e. g. \cite{H-W}, \cite{Gio},
\cite{E} which act in infinite dimensional Hilbert space.
Although many questions for Gaussian Bosonic systems with finite
number of modes can be solved with finite dimensional matrix techniques,
a general underlying Hilbert space operator analysis is indispensable.

Moreover, it was observed recently  \cite{Sh} that Shor's proof of
global equivalence of different forms of the famous additivity
conjecture is related to weird discontinuity of the
$\chi$-capacity in the  infinite dimensional case. All this calls
for a mathematically rigorous treatment involving specific results
from the operator theory in a Hilbert space and measure theory.

There are two important features essential for channels
in infinite dimensions. One is the necessity of the input constraints
(such as mean energy constraint for Gaussian channels) to prevent
from infinite capacities (although considering input constraints
was recently shown quite useful also in the study of the additivity
conjecture for channels in finite dimensions \cite{H-Sh}).
Another is the natural appearance of infinite, and, in general, ``continuous''
state ensembles understood as probability measures on the set of all
quantum states. By using compactness criteria from probability theory
and operator theory we can show that the set of all
generalized ensembles with the average  in a
compact set of states is itself a compact subset of the set of
all probability measures. With this in hand we give a sufficient
condition for existence of an
optimal generalized ensemble for a constrained quantum channel.
 This condition can be verified in particular in the case of Bosonic
Gaussian channels with constrained mean energy. In the case of
convex constraints we give a characterization of the optimal
generalized ensemble extending the ``maximal distance property''
\cite{Sch-West-1}, \cite{H-Sh}.

\section{Preliminaries}

Let $\mathcal{H}$ be a separable Hilbert space, $\mathfrak{B}(\mathcal{H})$
the algebra of all bounded operators in $\mathcal{H}$, $\mathfrak{T}(
\mathcal{H})$ the Banach space of all trace-class operators with
the trace norm $\Vert \cdot \Vert _{1}$ and
$\mathfrak{S}(\mathcal{H})$ the closed convex subset of
$\mathfrak{T}(\mathcal{H})$ consisting of all density operators
(states) in $\mathcal{H}$, which is complete separable metric
space with the metric defined by the norm. We shall use the fact that
convergence of a sequence of states to a \textit{state} in the
weak operator topology is equivalent to convergence of this
sequence  to this state in the trace norm \cite{D-A}.
A closed subset $\mathcal{K}$ of states is compact if and only if for
any $\varepsilon>0$ there is a finite dimensional projector $P$
such that $\mathrm{Tr}\rho P\geq 1-\varepsilon$ for all $\rho\in\mathcal{K}$
\cite{S}.

A finite collection $\{\pi _{i},\rho _{i}\}$ of states $\rho _{i}$
with the corresponding probabilities $\pi _{i}$ is conventionally
called \textit{ensemble}.
The state $\bar{\rho}=\sum_{i}\pi _{i}\rho _{i}$ is called \textit{the
average} of the ensemble.

We refer to \cite{Bil},\cite{Par} for definitions and facts
concerning probability measures on separable metric spaces.
In particular we denote $\mathrm{supp}(\pi)$ support of
measure $\pi$ as defined in \cite{Par}.

\textbf{Definition}. We call \textit{generalized ensemble} an
arbitrary Borel probability measure $\pi $ on
$\mathfrak{S}(\mathcal{H})$. The \textit{average}\footnote{Also
called barycenter of the measure $\pi$.} of the generalized
ensemble $\pi$ is defined by the  Pettis integral
\[
\bar{\rho}(\pi )=\int\limits_{\mathfrak{S}(\mathcal{H})}\rho \pi (d\rho ).
\]
Using the result of \cite{D-A} it is possible to show that the
above integral exists also in Bochner sense \cite{H&P}.

The conventional ensembles correspond to measures with finite support.

Denote by $\mathcal{P}$ the convex set
of all probability measures on $\mathfrak{S}(\mathcal{H})$
equipped with the topology of weak convergence \cite{Bil}. It is
easy to see (due to the result of \cite{D-A}) that the mapping
$\pi \mapsto \bar{\rho}(\pi )$ is continuous in this topology.

\textbf{Lemma 1.} \textit{The subset of measures with finite
support is dense in the set of all measures with given average
}$\bar{\rho}$\textit{.}

A proof of this statement is given in the appendix A.

In what follows $\log $ denotes the function on $[0,+\infty ),$ which
coincides with the usual logarithm on $\left( 0,+\infty \right) $ and
vanishes at zero. If $A$ is a positive finite rank operator in $\mathcal{H},$
then the entropy is defined as
\begin{equation}
H(A)=\mathrm{Tr}A\left( I\log \mathrm{Tr}A-\log A\right) ,
\label{ent}
\end{equation}
where $I$ is the unit operator in $\mathcal{H}$. If $A,B$ two such operators then the
relative entropy is defined as
\begin{equation}
H(A\,\Vert B)=\mathrm{Tr}(A\log A-A\log B+B-A)  \label{relent}
\end{equation}
provided $\mathrm{ran}A\subseteq\mathrm{ran}B$, and $H(A\,\Vert
B)=+\infty$ otherwise (throughout this paper $\mathrm{ran}$
denotes the closure of the range of an operator in $\mathcal{H}$).

These definitions can be extended to arbitrary positive $A$,$B\in
\mathfrak{T}(\mathcal{H})$ with the help of the following lemma
\cite{L}:

\textbf{Lemma 2.} \textit{Let $\left\{ P_{n}\right\} $ be an
arbitrary sequence of finite dimensional projectors monotonously
increasing to the
unit operator $I$. The sequences }$\left\{ H(P_{n}AP_{n})\right\} ,$ $
\left\{ H(P_{n}AP_{n}\Vert P_{n}BP_{n})\right\} $\textit{\ are monotonously
increasing and have the limits in the range }$\left[ 0,+\infty \right] $
\textit{\ independent of the choice of the sequence $\left\{ P_{n}\right\} .$
}

We can thus define the entropy and the relative entropy as\textit{\
\[
H(A)=\lim_{n\rightarrow +\infty }H(P_{n}AP_{n});\;\quad H(A\,\Vert
B)=\lim_{n\rightarrow +\infty }H(P_{n}AP_{n}\Vert P_{n}BP_{n}).
\]
}

As it is well known, the properties of the entropy for infinite and finite
dimensional Hilbert spaces differ quite substantially: in the latter case
the entropy is bounded continuous function on $\mathfrak{S}(\mathcal{H}),$ while
in the former it is discontinuous (lower semicontinuous) at every point, and
infinite  ``most everywhere'' in the sense
that the set of states with finite entropy is a first category subset of $
\mathfrak{S}(\mathcal{H})$ $\cite{W}.$

\section{The $\chi$-capacity of constrained channels}

Lemma 2 implies, in particular, that the nonnegative
function\break $\rho\mapsto H(\Phi (\rho )\Vert \Phi
(\bar{\rho}(\pi )))$ is measurable on $
\mathfrak{S}(\mathcal{H})$. Hence the functional
\[
\chi _{\Phi }(\pi )=\int\limits_{\mathfrak{S}(\mathcal{H})}H(\Phi (\rho
)\Vert \Phi (\bar{\rho}(\pi )))\pi (d\rho )
\]
is well defined on the set
$\mathcal{P}$ (with the range
$[0;+\infty ]$).

\textbf{Proposition 1.} \textit{The functional $\chi _{\Phi }(\pi
)$ is lower semicontinuous on $\mathcal{P}$. If $H(\Phi
(\bar{\rho}(\pi ))<\infty,$ then}
\begin{equation}\label{formula}
\chi _{\Phi }(\pi)= H(\Phi (\bar{\rho}(\pi
)))-\int\limits_{\mathfrak{S}( \mathcal{H})}H(\Phi (\rho ))\pi
(d\rho ).
\end{equation}

\textbf{Proof.} Let $\left\{ P_{n}\right\} $ be an
arbitrary sequence of finite dimensional projectors monotonously
increasing to the
unit operator $I$. We show first that the functionals
\[
\chi _{\Phi }^{n}(\pi )=\int\limits_{\mathfrak{S}(\mathcal{H})}H(P_{n}\Phi
(\rho )P_{n}\Vert P_{n}\Phi (\bar{\rho}(\pi ))P_{n})\pi (d\rho )
\]
are continuous.

We have
\[
\mathrm{ran}(P_{n}\Phi (\rho )P_{n})\subseteq \mathrm{ran}(P_{n}\Phi (\bar{
\rho}(\pi ))P_{n})
\]
for $\pi -$almost all $\rho $. Indeed, closure of the range is orthogonal
complement to the null subspace of a Hermitian operator, and for null
subspaces the opposite inclusion holds obviously. It follows that
\[
\begin{array}{c}
H(P_{n}\Phi (\rho )P_{n}\Vert P_{n}\Phi (\bar{\rho}(\pi
))P_{n})=\mathrm{Tr}((P_{n}\Phi (\rho )P_{n})\log (P_{n}\Phi (\rho
)P_{n}) \\\\ -(P_{n}\Phi(\rho)P_{n})\log (P_{n}\Phi
(\bar{\rho}(\pi ))P_{n})+P_{n}\Phi (\bar{\rho}(\pi
))P_{n}-P_{n}\Phi (\rho )P_{n})
\end{array}
\]
for $\pi -$almost all $\rho $. By using (\ref{ent}) we have
\[
\begin{array}{c}
\chi _{\Phi }^{n}(\pi )=-\int\limits_{\mathfrak{S}(\mathcal{H}
)}H(P_{n}\Phi (\rho )P_{n})\pi (d\rho )+
\int\limits_{\mathfrak{S}(\mathcal{H})}\mathrm{Tr}(P_{n}\Phi (\rho
))\log \mathrm{Tr}(P_{n}\Phi (\rho ))\pi (d\rho ) \\\\
-\int\limits_{\mathfrak{S}(\mathcal{H})}\mathrm{Tr}(P_{n}\Phi
(\rho )P_{n})\log (P_{n}\Phi (\bar{\rho}(\pi ))P_{n})\pi
(d\rho)\\\\
+\int\limits_{\mathfrak{S}(\mathcal{H})}\mathrm{Tr}(P_{n}\Phi (\bar{\rho}
(\pi )))\pi (d\rho )-\int\limits_{\mathfrak{S}(\mathcal{H})}\mathrm{Tr}
(P_{n}\Phi (\rho))\pi (d\rho ).
\end{array}
\]
It is easy to see that the two last terms cancel while the central
term can be transformed in the following way
\[
\begin{array}{c}
-\int\limits_{\mathfrak{S}(\mathcal{H})}\mathrm{Tr}(P_{n}\Phi
(\rho )P_{n})\log (P_{n}\Phi (\bar{\rho}(\pi ))P_{n})\pi (d\rho )
\\\\ =-\mathrm{Tr}\int\limits_{\mathfrak{S}(\mathcal{H})}
(P_{n}\Phi (\rho )P_{n})\log (P_{n}\Phi (\bar{\rho}(\pi
))P_{n})\pi (d\rho )\\\\ =H(P_{n}\Phi (\bar{\rho}(\pi
))P_{n})-\mathrm{Tr}(P_{n}\Phi (\bar{\rho}(\pi )))\log
\mathrm{Tr}(P_{n}\Phi (\bar{\rho}(\pi ))).
\end{array}
\]
Hence
\begin{equation}\label{chi-n}
\begin{array}{c}
\chi _{\Phi }^{n}(\pi )=H(P_{n}\Phi (\bar{\rho}(\pi
))P_{n})-\mathrm{Tr}(P_{n}\Phi (\bar{\rho}(\pi )))\log
\mathrm{Tr}(P_{n}\Phi (\bar{\rho}(\pi
)))\\\\-\int\limits_{\mathfrak{S}(\mathcal{H})}H(P_{n}\Phi(\rho
)P_{n})\pi (d\rho
)+\!\!\!\int\limits_{\mathfrak{S}(\mathcal{H})}\!\mathrm{Tr}(P_{n}\Phi
(\rho ))\log \mathrm{Tr}(P_{n}\Phi (\rho ))\pi (d\rho ).
\end{array}
\end{equation}
Continuity and boundedness of the quantum entropy in the finite
dimensional case and similar properties of the function
$\rho\mapsto\mathrm{Tr}(P_{n}\Phi (\rho ))\log
\mathrm{Tr}(P_{n}\Phi (\rho ))$ imply continuity of the
functionals $\chi _{\Phi }^{n}(\pi ).$

By the monotonous convergence theorem (in what follows,
m.c.-theorem) \cite{K&F},\cite{H&P} the sequence of functionals
$\chi _{\Phi }^{n}(\pi )$ is monotonously increasing and pointwise
converges to $\chi _{\Phi }(\pi )$. Hence the functional
$\chi_{\Phi }(\pi )$ is lower semicontinuous.

To prove (\ref{formula}) note that lemma 2 implies
$$
\lim_{n\rightarrow+\infty}H(P_{n}\Phi (\bar{\rho}(\pi
))P_{n})=H(\Phi (\bar{\rho}(\pi )))
$$
and
$$
\lim_{n\rightarrow+\infty}\int\limits_{\mathfrak{S}(\mathcal{H})}H(P_{n}\Phi(\rho
)P_{n})\pi (d\rho
)=\int\limits_{\mathfrak{S}(\mathcal{H})}H(\Phi(\rho))\pi (d\rho)
$$
due to m.c.-theorem. For every $\rho$ the sequence $\{\mathrm{Tr}(P_{n}\Phi (\rho))\}$
is in $[0,1]$ and converges to $1$, therefore
$
\lim_{n\rightarrow+\infty}\mathrm{Tr}(P_{n}(\rho))\log \mathrm{Tr}(P_{n}(\rho))=0,
$
in particular the second term in (\ref{chi-n}) tends to $0$.
Since $|x\log x|<1$ for all
$x\in(0,1],$ the last term also tends to $0$
by dominated convergence theorem, so passing to the limit $n\to\infty$ in (\ref{chi-n})
gives (\ref{formula}).$\square$

Let $\mathcal{H},\mathcal{H}^{\prime }$ be a pair of separable Hilbert
spaces which we shall call correspondingly input and output space. A channel
$\Phi $ is a linear positive trace preserving map from $\mathfrak{T}(\mathcal{
H })$ to $\mathfrak{T}(\mathcal{H}^{\prime })$ such that the dual
map $\Phi ^{\ast }:\mathfrak{B}(\mathcal{H}^{\prime
})\mapsto\mathfrak{B}(\mathcal{H})$ (which exists since $\Phi $ is
bounded) is completely positive. Let $\mathcal{A}$ be an arbitrary
subset of $ \mathfrak{S}(\mathcal{H})$. We consider
constraint on input ensemble $\{\pi _{i},\rho _{i}\}$, defined by
the requirement $\bar{\rho}\in \mathcal{A}$. The channel $\Phi $
with this constraint is called the $\mathcal{A}$ -\textit{constrained}
channel. We define the $\chi $-\textit{capacity} of the
$\mathcal{A}$-constrained channel $\Phi $ as
\begin{equation}
\bar{C}(\Phi ;\mathcal{A})=\sup_{\bar{\rho}\in \mathcal{A}}\chi _{\Phi
}(\{\pi _{i},\rho _{i}\}),  \label{ccap}
\end{equation}
where
\begin{equation}
\chi _{\Phi }(\{\pi _{i},\rho _{i}\})=\sum_{i}\pi _{i}H(\Phi (\rho
_{i})\Vert \Phi (\bar{\rho})).  \label{h-q}
\end{equation}

\textit{Throughout this paper we shall consider the constraint sets $
\mathcal{A}$ such that}
\begin{equation}
\bar{C}(\Phi ;\mathcal{A})<+\infty .  \label{A}
\end{equation}

The subset of $\mathcal{P}$,
consisting of all measures $\pi $ with the average state
$\bar{\rho}(\pi)$ in a subset $ \mathcal{A\subseteq
\mathfrak{S}(\mathcal{H})}$, will be denoted $\mathcal{P}
_{\mathcal{A}}$. Lemma 1 and proposition 1 imply

\textbf{Corollary 1.} \textit{The $\chi$-capacity of
$\mathcal{A}$-constrained channel $\Phi$ can be defined by}
\[
\bar{C}(\Phi ;\mathcal{A})=\!\!\sup\limits_{\pi \in \mathcal{P}_{\mathcal{A}
}}\chi _{\Phi }(\pi ).
\]
\textbf{Proof.} The definition (\ref{ccap}) is a similar expression
in which the supremum is over all measures in
$\mathcal{P}_{\mathcal{A}}$ with finite support. By lemma 1 we can
approximate arbitrary measure $\pi$ in $\mathcal{P}_{\mathcal{A}}$
by a sequence $\{\pi_{n}\}$ of measures in
$\mathcal{P}_{\mathcal{A}}$ with finite support. By proposition 1,
$\liminf_{n\rightarrow+\infty}\chi _{\Phi }(\pi_{n})\geq\chi
_{\Phi }(\pi)$. It follows that the supremum over all measures in
$\mathcal{P}_{\mathcal{A}}$ coincides with the supremum over all
measures in $\mathcal{P}_{\mathcal{A}}$ with finite
support.$\square$

\section{Compact constraints}
It is convenient to introduce the following notion. An
unbounded positive operator $H$ in $\mathcal{H}$ with
discrete spectrum of finite multiplicity will be called
$\mathfrak{H}$-\textit{operator}. Let $Q_n$ be the spectral projector
of $H$ corresponding to the lowest $n$ eigenvalues. Following \cite{H-c-w-c}
we shall denote
\begin{equation}\label{limit}\mathrm{Tr}\rho H=\lim_{n\to\infty}\mathrm{Tr}\rho Q_n H,
\end{equation}
where the sequence on the right side is monotonously nondecreasing.
It was shown in \cite{H-c-w-c} that
\begin{equation}
\mathcal{K}=\left\{ \rho :\mathrm{Tr}\rho H\leq h\right\} ,  \label{ham}
\end{equation}
where $H$ is an $\mathfrak{H}$-operator, is a compact subset of
\textit{$\mathfrak{S}(\mathcal{H})$ }.

\textbf{Lemma 3.} \textit{Let $\mathcal{A}$ be a compact subset of $
\mathfrak{S}(\mathcal{H})$. Then there exist an
$\mathfrak{H}$-operator $H$ and a positive number $h$ such that
$\mathrm{Tr}\rho H\leq h$ for all
$\rho\in\mathcal{A}$.}

\textbf{Proof.} By the compactness criterion from \cite{S} for any
natural $n$ there exists a finite rank projector $P_{n}$ such that
$\mathrm{Tr}\rho P_{n}\geq 1-n^{-3}$ for all $\rho $ in
$\mathcal{A}$. Without loss of generality we may assume that
$\bigvee_{k=1}^{+\infty}P_{k}(\mathcal{H})=\mathcal{H}$, where
$\bigvee$ denotes closed linear span of the subspaces. Let
$\hat{P}_{n}$ be the
projector on the finite dimensional subspace $\bigvee_{k=1}^{n}P_{k}(\mathcal{H
})$. Thus
$H=\sum_{n=1}^{+\infty }n(\hat{P}_{n+1}-\hat{P}_{n})$ is a $\mathfrak{H}$-operator
satisfying
\[
\mathrm{Tr}\rho H=\sum_{n=1}^{+\infty }n\mathrm{Tr}\rho (\hat{P}_{n+1}-\hat{P
}_{n})\leq \sum_{n=1}^{+\infty }n\mathrm{Tr}\rho (I_{\mathcal{H}}-\hat{P}
_{n})\leq \sum_{n=1}^{+\infty }n^{-2}=h
\]
for arbitrary state $\rho $ in the set $\mathcal{A}$. $\square $

\textbf{Proposition 2.} \textit{The set
$\mathcal{P}_{\mathcal{A}}$ is a compact subset of
$\mathcal{P}$ if and only if the set
$\mathcal{A}$ is a compact subset of $\mathfrak{S}(\mathcal{H})$.}

\textbf{Proof.} Let the set $\mathcal{P}_{\mathcal{A}}$ be
compact. The set $\mathcal{A}$ is the image of the set
$\mathcal{P}_{ \mathcal{A}}$ under the continuous mapping $\pi
\mapsto \bar{\rho}(\pi )$, hence it is compact.

Let the set $\mathcal{A}$ be compact. By lemma 3 there exists an
$\mathfrak{H}$-operator $H$ such that $\mathrm{Tr}\rho H\leq h$
for all $\rho $ in $\mathcal{A}$. For arbitrary $\pi \in
\mathcal{P}_{\mathcal{A}}$ we have

\begin{equation}
\int\limits_{\mathfrak{S}(\mathcal{H})}(\mathrm{Tr}\rho H) \pi (d\rho
)=\mathrm{Tr}\left(\;\int\limits_{\mathfrak{S}(\mathcal{H})}\rho
\pi (d\rho)\;H\right) =\mathrm{Tr}\bar{\rho}(\pi)H\leq h
\label{int-eq}
\end{equation}
The existence of the integral on the left side and the first
equality follows from m.c.-theorem, since by (\ref{limit}) the
function $\mathrm{Tr}\rho H$ is the limit of nondecreasing sequence
of continuous bounded functions $\mathrm{Tr}\rho Q_{n}H$.

Let $\mathcal{K}_{\varepsilon }=\{\rho :\mathrm{Tr}\rho H\leq
h\varepsilon ^{-1}\}$. The set $\mathcal{K}_{\varepsilon }$ is a
compact subset of $ \mathfrak{S}(\mathcal{H})$ for any
$\varepsilon $. By (\ref{int-eq}) for any measure $\pi $ in
$\mathcal{P}_{\mathcal{A}}$ we have
\begin{equation}
\begin{array}{c}
\pi (\mathfrak{S}(\mathcal{H})\backslash \mathcal{K}_{\varepsilon
})=\int\limits_{\mathfrak{S}(\mathcal{H})\backslash
\mathcal{K}_{\varepsilon }}\pi (d\rho )\leq \varepsilon
h^{-1}\int\limits_{\mathfrak{S}(\mathcal{H} )\backslash
\mathcal{K}_{\varepsilon }}(\mathrm{Tr}\rho H) \pi (d\rho )\leq
\varepsilon
\end{array}
\label{int-ineq}
\end{equation}
By Prokhorov's theorem \cite{P} the (obviously closed) set
$\mathcal{P}_{\mathcal{A}}$ is compact.$ \square $

We will use the following notions, introduced in \cite{Sh}. The sequence of
ensembles $\{\pi _{i}^{k},\rho _{i}^{k}\}$ with the averages $\,\bar{\rho}
^{k}\in \mathcal{A}$ is called an \textit{approximating sequence} if
\[
\lim_{k\rightarrow +\infty }\chi _{\Phi }(\{\pi _{i}^{k},\rho _{i}^{k}\})=
\bar{C}(\Phi ;\mathcal{A}).
\]
The state $\bar{\rho}\in \mathcal{A}$ is called an \textit{optimal
average state } if it is a partial limit of a sequence of average
states for some approximating sequence of ensembles. Compactness of
the set $\mathcal{A}$ implies that the set of optimal average
states is not empty.

\textbf{Theorem.} \textit{If the restriction of the output entropy
$H(\Phi(\rho))$ to the set $\mathcal{A}$ is continuous at least at
one optimal average state $\bar{\rho}_{0}\in\mathcal{A}$ then
there exist an optimal generalized ensemble $\pi^{*}$ in
$\mathcal{P}_{\mathcal{A}}$ such that
$\mathrm{supp}\pi^{*}\subseteq\mathrm{Extr}\mathfrak{S}(\mathcal{H})$
and
$$
\bar{C}(\Phi
;\mathcal{A})=\chi_{\Phi}(\pi^{*})=
\int\limits_{\mathfrak{S}(\mathcal{H})}H(\Phi(\rho)\|\Phi(\bar{\rho}(\pi^{*})))\pi^{*}(d\rho).
$$}

\textbf{Proof.} We will show first that the function
\[
\pi \mapsto \int\limits_{\mathfrak{S}(\mathcal{H})}H(\Phi (\rho ))\pi (d\rho
)
\]
is well defined and lower semicontinuous on the set $\mathcal{P}_{
\mathcal{A}}$.

By lemma 2 the function $H(\Phi (\rho ))$ is a pointwise limit of the
monotonously increasing sequences of functions
\[
f_{n}(\rho )=\mathrm{Tr}\left((P_{n}\Phi (\rho )P_{n})\left( I\log \mathrm{Tr}
(P_{n}\Phi (\rho )P_{n})-\log (P_{n}\Phi (\rho
)P_{n})\right)\right) ,
\]
which are continuous and bounded on $\mathfrak{S}(\mathcal{H})$.
Hence the function $H(\Phi (\rho ))$ is measurable and the
m.c.-theorem implies
\[
\int\limits_{\mathfrak{S}(\mathcal{H})}H(\Phi (\rho ))\pi (d\rho
)=\lim_{n\rightarrow \infty }\int\limits_{\mathfrak{S}(\mathcal{H}
)}f_{n}(\rho )\pi (d\rho ).
\]
The sequence of continuous functionals
\[
\pi \mapsto \int\limits_{\mathfrak{S}(\mathcal{H})}f_{n}(\rho )\pi (d\rho )
\]
is nondecreasing. Hence its pointwise limit is lower semicontinuous.

By the assumption the restriction of the function $H(\Phi (\rho
))$ to the set $\mathcal{A}$ is continuous at some optimal average
state $\bar{\rho}_{0} $. The continuity of the mapping $\pi
\mapsto \bar{\rho}(\pi )$ implies that the restriction of the
functional $\pi \mapsto H(\Phi (\bar{\rho}(\pi )))$ to the set
$\mathcal{P}_{\mathcal{A}}$ is continuous at any point $\pi _{0}$
such that $\bar{\rho}(\pi _{0})=\bar{\rho}_{0}$. Hence
$H(\Phi(\bar{\rho}(\pi)))<+\infty$ for any point $\pi$ in the
intersection of $\mathcal{P}_{\mathcal{A}}$ with some
neighbourhood of $\pi_{0}$.  For every such point $\pi$ the
relation (\ref{formula}) holds. Therefore the restriction of the
functional $\chi _{\Phi }(\pi)$ to the set
$\mathcal{P}_{\mathcal{A}}$ is upper semicontinuous, and by
proposition 1 it is continuous at any point $\pi _{0}$ in
$\mathcal{P}_{\mathcal{A}}$ such that $\bar{\rho}(\pi
_{0})=\bar{\rho}_{0}$.

Let $\{\pi _{i}^{n},\rho _{i}^{n}\}$ be an approximating sequence
of ensembles with the corresponding sequence of average states
$\bar{\rho}^{n}$ converging to the state $\bar{\rho}_{0}$.
Decomposing each state of the ensemble $\{\pi _{i}^{n},\rho
_{i}^{n}\}$ into a countable convex combination of pure states we
obtain the sequence $\{\hat{\pi}_{j}^{n},\hat{\rho}_{j}^{n}\}$ of
generalized ensembles consisting of countable number of pure
states with the same sequence of the average states
$\bar{\rho}^{n}$. Let $\hat{\pi}^{n}$ be
the sequence of measures ascribing value $\hat{\pi}_{j}^{n}$ to the set $\{
\hat{\rho}_{j}^{n}\}$ for each $j$. It follows that
\begin{equation}\label{chi-exp}
\chi _{\Phi
}(\hat{\pi}_{n})=\sum\limits_{j}\hat{\pi}_{j}^{n}H(\Phi
(\hat{\rho}_{j}^{n})\Vert \Phi (\bar{\rho}^{n}))\geq
\sum\limits_{i}\pi _{i}^{n}H(\Phi (\rho _{i}^{n})\Vert \Phi
(\bar{\rho}^{n}))=\chi _{\Phi }(\{\pi _{i}^{n},\rho _{i}^{n}\}),
\end{equation}
where the inequality follows from convexity of the relative entropy. By
construction $\mathrm{supp}\hat{\pi}^{n}\subseteq \mathrm{Extr}\mathfrak{S}(
\mathcal{H})$ for each $n$. By proposition 2 there exists a subsequence $
\hat{\pi}^{n_{k}}$, converging to some measure $\pi ^{\ast }$ in $\mathcal{P}
_{\mathcal{A}}$. Since the set $\mathrm{Extr}\mathfrak{S}(\mathcal{H})$ of
all pure states is closed subset of $\mathfrak{S}(\mathcal{H})$\footnote{
The set $\mathrm{Extr}\mathfrak{S}(\mathcal{H})$ is described by the
inequality $H(\rho )\leq 0$, and due to lower semicontinuity of the quantum
entropy it is closed.} we have $\mathrm{supp}\pi ^{\ast }\subseteq \mathrm{
Extr}\mathfrak{S}(\mathcal{H})$ due to theorem 6.1 in \cite{Par}.
It is clear that $\bar{\rho}(\pi ^{\ast })=\bar{\rho}_{0}$ and,
hence, as shown above, the restriction of the functional
$\chi _{\Phi }(\pi)$ on the set $\mathcal{P}_{\mathcal{A}}$ is
continuous at the point $\pi ^{\ast }$. This, the approximating
property of the sequence $\{\pi _{i}^{n},\rho _{i}^{n}\}$ and
(\ref{chi-exp}) implies
\[
\bar{C}(\Phi ;\mathcal{A})=\lim_{k\rightarrow \infty }\chi _{\Phi }(\{\pi
_{i}^{n_{k}},\rho _{i}^{n_{k}}\})\leq \lim_{k\rightarrow \infty }\chi _{\Phi
}(\hat{\pi}_{n_{k}})=\chi _{\Phi }(\pi ^{\ast }).
\]
Since the converse inequality follows from corollary 1, we obtain $\bar{C}
(\Phi ;\mathcal{A})=\chi _{\Phi }(\pi ^{\ast })$, which means that
the measure  $\pi ^{\ast }$ is an optimal generalized ensemble for
the $\mathcal{A}$-constrained channel $\Phi $.$\square $

\bigskip \textbf{Corollary 2.} \textit{For arbitrary state $\rho _{0}$ with $
H(\Phi (\rho _{0}))<+\infty $ there exists a generalized
ensemble\footnote{In what follows we can consider the
generalized ensembles as measures supported by the set of pure states}
$\pi _{0}$ such that $
\bar{\rho}(\pi _{0})=\rho _{0}$ and
\[
\chi _{\Phi }(\rho _{0})\equiv \sup_{\sum_{i}\pi _{i}\rho _{i}=\rho
_{0}}\chi _{\Phi }(\{\pi _{i},\rho _{i}\})=\int\limits_{\mathfrak{S}(
\mathcal{H})}H(\Phi (\rho )\Vert \Phi (\rho _{0}))\pi _{0}(d\rho ).
\]
} \textbf{Proof.} It is sufficient to note that the condition of
the theorem holds trivially for $\mathcal{A}=\{\rho _{0}\}$.
$\square$

In the finite dimensional case we obviously have
\begin{equation}\label{cap-chi-rel}
\bar{C}(\Phi ;\mathcal{A})=\chi _{\Phi }(\bar{\rho}),
\end{equation}
where $\bar{\rho}$ is the average state of any optimal ensemble.
The generalization of this relation to the infinite dimensional
case is closely connected with the question of existence of the
optimal generalized ensemble.

\bigskip \textbf{Corollary 3.} \textit{If an optimal generalized ensemble
for the $\mathcal{A}$ - constrained channel $\Phi$ exists, then the equality
(\ref{cap-chi-rel}) holds for some optimal average state
$\bar{\rho}$ for the $\mathcal{A}$-constrained channel $\Phi$.}

\textit{If the equality (\ref{cap-chi-rel}) holds for some optimal
average state $\bar{\rho}$ for the $\mathcal{A}$ - constrained
channel $\Phi$ with $H(\Phi (\bar{\rho}))<+\infty$ then there
exists an optimal generalized ensemble for the
$\mathcal{A}$-constrained channel $\Phi$.}

\textbf{Proof.} The first assertion is obvious while the second
one follows from corollary 2.$\square$

\textbf{Remark.} The continuity condition in the theorem
is essential, as it is shown in Appendix B. It is possible to show that this condition
 holds automatically if the set $\mathcal{A}$ is convex with a
finite number of extreme points with finite output entropy. We
conjecture that this condition holds for arbitrary convex compact
set $\mathcal{A}$ due to the special properties of optimal average
states in this case, considered in \cite{Sh}.

\textbf{Proposition 3.} \textit{Let $H^{\prime }$ be a
$\mathfrak{H}$-operator on the space $\mathcal{H}^{\prime }$ such
that }
\begin{equation}
\mathit{\mathrm{Tr}\exp (-\beta H^{\prime })<+\infty \quad \ \mathrm{for\
all }\ \quad \beta >0}  \label{beta}
\end{equation}
\textit{and $\mathrm{Tr}\,\Phi (\rho)H^{\prime }\leq h^{\prime }$
for all $\rho\in\mathcal{A}$. Then there exists an optimal
generalized ensemble for the $\mathcal{A}$-constrained channel
$\Phi $.}

\textbf{Proof.} We will show that under the condition of the lemma
the restriction of the output entropy $H(\Phi (\rho ))$ on the set
$\mathcal{A}$ is continuous, which implies validity of the
condition of the theorem.

Let $\rho _{\beta }^{\prime }=(\mathrm{Tr}\exp (-\beta H^{\prime
}))^{-1}\exp (-\beta H^{\prime })$ be a state in $\mathfrak{S}(\mathcal{H}
^{\prime })$. For arbitrary $\rho $ in $\mathcal{A}$ we have
\begin{equation}
H(\Phi (\rho )\Vert \rho _{\beta }^{\prime })=-H(\Phi (\rho ))+\beta \mathrm{
Tr}\Phi (\rho )H^{\prime }+\log \mathrm{Tr}\exp (-\beta H^{\prime })
\label{r-e-exp}
\end{equation}
Let $\rho _{n}$ be an arbitrary sequence of states in $\mathcal{A}$
converging to the state $\rho $. By using (\ref{r-e-exp}) and lower
semicontinuity of the relative entropy we obtain
\[
\begin{array}{c}
\limsup\limits_{n\rightarrow \infty }H(\Phi (\rho _{n}))=H(\Phi (\rho
))+H(\Phi (\rho )\Vert \rho _{\beta }^{\prime
})-\liminf\limits_{n\rightarrow \infty }H(\Phi (\rho _{n})\Vert \rho _{\beta
}^{\prime }) \\
+\limsup\limits_{n\rightarrow \infty }\beta \mathrm{Tr}\Phi (\rho
_{n})H^{\prime }-\beta \mathrm{Tr}\Phi (\rho )H^{\prime }\leq H(\Phi (\rho
))+\beta h^{\prime }.
\end{array}
\]
By tending $\beta $ in the above inequality to zero we can
establish the upper semicontinuity of the restriction of the
function $H(\Phi (\rho ))$ to the set $\mathcal{A}$. The lower
semicontinuity of this function follows from the lower
semicontinuity of the entropy \cite{W}. Hence the restriction of
the function $H(\Phi (\rho ))$ on the set $\mathcal{A}$ is
continuous. $\square $

The condition of proposition 3 is fulfilled for Gaussian
channels with the power constraint of the form (\ref{ham}) where
$H=R^{T}\epsilon R$ is
the many-mode oscillator Hamiltonian with nondegenerate energy matrix $
\epsilon ,$ and $R$ are the canonical variables of the system. We give a
brief sketch of the argument which can be made rigorous by taking care of
unboundedness of the canonical variables. Indeed, let
\[
R^{\prime }=KR+K_{E}R_{E}
\]
be the equation of the channel in the Heisenberg picture, where $R_{E}$ are
the canonical variables of the environment which is in the Gaussian state
with zero mean and the correlation matrix $\alpha _{E}$ \cite{H-W}. Taking $
H^{\prime }=c[R^{T}R-I$Sp$\alpha _{E}K_{E}^{T}$ $K_{E}]$, we have $\Phi
^{\ast }(H^{\prime })=cR^{T}K^{T}KR$, and we can always choose a positive $c$
such that $\Phi ^{\ast }(H^{\prime })\leq H.$ Moreover, $H^{\prime }$
satisfies the condition (\ref{beta}). Thus the conditions of proposition 3
can be fulfilled in this case.

\textbf{Conjecture.} \textit{For arbitrary Gaussian channel with
the power constraint an optimal generalized ensemble is given by a
Gaussian measure supported by the set of pure Gaussian states with
arbitrary mean and a fixed correlation matrix.}

This conjecture was stated in \cite{H-W} for
attenuation/amplification channel with classical noise. For the
case of pure attenuation channel characterized by the property of
zero minimal output entropy the validity of this conjecture was
established in \cite{Gio}.

\section{Convex constraints}
In the case of convex constraint set there are further special
properties, such as uniqueness of the output of optimal average
state, see \cite{Sh}. The following lemma is a generalization of
Donald's identity \cite{Don}.

\textbf{Lemma 4.} \textit{For arbitrary measure $\pi$ in
$\mathcal{P}$ and arbitrary state
$\sigma$ in $\mathfrak{S}(\mathcal{H})$ the following identity
holds
\begin{equation}
\int\limits_{\mathfrak{S}(\mathcal{H})}H(\rho \Vert \sigma
)\pi(d\rho )=\int\limits_{\mathfrak{S}(\mathcal{H})}H(\rho \Vert
\bar{\rho}(\pi ))\pi (d\rho )+H(\bar{\rho}(\pi )\Vert \sigma ).
\label{relens}
\end{equation}
}

\textbf{Proof.} We first notice that in the finite dimensional
case Donald's identity
\[
\sum_{i}\pi _{i}H(\rho_{i}\Vert \sigma )=\sum_{i}\pi _{i}H(\rho_{i}\Vert \bar{\rho}
(\pi ))+H(\bar{\rho}(\pi )\Vert \sigma )
\]
holds for not necessarily normalized positive operators with the
generalized definition of the relative entropy (\ref{relent}).
This can be obviously extended to generalized ensembles in
finite-dimensional Hilbert space, giving (\ref{relens}) for this
case. Thus this relation holds for the operators $P_{n}\rho
P_{n},P_{n}\sigma P_{n}$, where $P_{n}$ is an arbitrary sequence
of finite projectors increasing to $I_{\mathcal{H}}$. Passing to
the limit $n\rightarrow \infty $ and referring to the m.c.-theorem,
we obtain (\ref{relens}) in infinite-dimensional case. $\square $

The following proposition is a generalization of the ``maximal distance
property'', cf. proposition 1 in
\cite{H-Sh}.

\textbf{Proposition 4.} \textit{Let $\mathcal{A}$ be convex subset
of $\mathfrak{S}(\mathcal{H})$. A measure $\pi \in
\mathcal{P}_{\mathcal{A}}$
is optimal generalized ensemble for the $\mathcal{A}$-constrained channel $
\Phi $ if and only if
\begin{equation}
\int\limits_{\mathfrak{S}(\mathcal{H})}H(\Phi (\rho )\Vert \Phi (\bar{\rho}
(\pi )))\mu (d\rho )\leq \int\limits_{\mathfrak{S}(\mathcal{H})}H(\Phi (\rho
)\Vert \Phi (\bar{\rho}(\pi )))\pi (d\rho )=\chi _{\Phi }(\pi )
\label{opt-ch}
\end{equation}
for arbitrary measure $\mu \in \mathcal{P}_{\mathcal{A}}$.}

\textbf{Proof.} Let inequality (\ref{opt-ch}) holds for arbitrary measure $
\mu \in \mathcal{P}_{\mathcal{A}}$. By lemma 4 we have
\[
\begin{array}{c}
\chi _{\Phi }(\mu )\leq \int\limits_{\mathfrak{S}(\mathcal{H})}H(\Phi (\rho
)\Vert \Phi (\bar{\rho}(\mu )))\mu (d\rho )+H(\Phi (\bar{\rho}(\mu ))\Vert
\Phi (\bar{\rho}(\pi ))) \\
=\int\limits_{\mathfrak{S}(\mathcal{H})}H(\Phi (\rho )\Vert \Phi (\bar{\rho}
(\pi )))\mu (d\rho )\leq \chi _{\Phi }(\pi ),
\end{array}
\]
which implies optimality of the measure $\pi $.

Conversely, let $\pi $ be an optimal generalized ensemble for the
$\mathcal{A}$-constrained channel $\Phi $ and $\mu $ be an
arbitrary measure in $\mathcal{P}_{\mathcal{A}}$. By convexity of
the set $\mathcal{A}$ the measure $\pi _{\eta }=\eta \mu +(1-\eta
)\pi $ is also in $\mathcal{P}_{\mathcal{A}}$ for arbitrary $\eta
\in (0,1)$. Using lemma 4 we have
\[
\begin{array}{c}
\chi _{\Phi }(\pi _{\eta })=\int\limits_{\mathfrak{S}(\mathcal{H})}H(\Phi
(\rho )\Vert \Phi (\bar{\rho}(\pi _{\eta })))\pi _{\eta }(d\rho ) \\
=\eta \int\limits_{\mathfrak{S}(\mathcal{H})}H(\Phi (\rho )\Vert \Phi (\bar{
\rho}(\pi _{\eta })))\mu (d\rho )+(1-\eta )\chi _{\Phi }(\pi )+(1-\eta )H(
\bar{\rho}(\pi )\Vert \bar{\rho}(\pi _{\eta })).
\end{array}
\]
The optimality of $\pi $ and nonnegativity of the relative entropy imply
\begin{equation}
\begin{array}{c}
\int\limits_{\mathfrak{S}(\mathcal{H})}H(\Phi (\rho ))\Vert \Phi (\bar{\rho}
(\pi _{\eta })))\mu (d\rho )-\chi _{\Phi }(\pi )\leq \eta ^{-1}(\chi _{\Phi
}(\pi _{\eta })-\chi _{\Phi }(\pi ))\leq 0.
\end{array}
\label{chi-ineq-one}
\end{equation}
By lemma 4 and lower semicontinuity of the relative entropy
\[
\begin{array}{c}
\liminf\limits_{\eta \rightarrow 0}\int\limits_{\mathfrak{S}(\mathcal{H}
)}H(\Phi (\rho ))\Vert \Phi (\bar{\rho}(\pi _{\eta })))\mu (d\rho ) \\
=\int\limits_{\mathfrak{S}(\mathcal{H})}H(\Phi (\rho ))\Vert \Phi (\bar{\rho}
(\mu )))\mu (d\rho )+\liminf\limits_{\eta \rightarrow 0}H(\Phi (\bar{\rho}
(\mu ))\Vert \bar{\rho}(\pi _{\eta })) \\
\geq \int\limits_{\mathfrak{S}(\mathcal{H})}H(\Phi (\rho ))\Vert \Phi (\bar{
\rho}(\mu )))\mu (d\rho )+H(\Phi (\bar{\rho}(\mu ))\Vert \bar{\rho}(\pi ))
\\
=\int\limits_{\mathfrak{S}(\mathcal{H})}H(\Phi (\rho ))\Vert \Phi (\bar{\rho}
(\pi )))\mu (d\rho )
\end{array}
\]
Then (\ref{chi-ineq-one}) implies
\[
\begin{array}{c}
\int\limits_{\mathfrak{S}(\mathcal{H})}H(\Phi (\rho ))\Vert \Phi (\bar{\rho}
(\pi )))\mu (d\rho )-\chi _{\Phi }(\pi ) \\
\leq \liminf\limits_{\eta \rightarrow 0}\int\limits_{\mathfrak{S}(\mathcal{H}
)}H(\Phi (\rho ))\Vert \Phi (\bar{\rho}(\pi _{\eta })))\mu (d\rho )-\chi
_{\Phi }(\pi ) \\
\leq \liminf\limits_{\eta \rightarrow 0}\eta ^{-1}(\chi _{\Phi }(\pi _{\eta
})-\chi _{\Phi }(\pi ))\leq 0.\;\square
\end{array}
\]

\section{Appendices}

\textbf{A. Proof of lemma 1.} We first notice that
$\mathrm{supp}(\pi )\subseteq U$, where $U$ a closed convex
subset of $\mathfrak{S}(\mathcal{H})$\textit{\ }implies
\begin{equation}
\bar{\rho}(\pi )\in U.  \label{l1}
\end{equation}
\textit{\ }This is obvious for arbitrary measure $\pi $ with
finite support. By theorem 6.3 in \cite{Par} the set of such
measures is dense in $\mathcal{P}$.
The continuity of the mapping $\pi \mapsto \bar{\rho}(\pi )$
completes the proof of (\ref{l1}).

Let now $\pi $ be an arbitrary measure in
$\mathcal{P}$. Since
$\mathfrak{S}(\mathcal{H})$ is separable we can, for each $n\in
\mathbb{N}$, find a sequence $\left\{ A_{i}^{n}\right\} $ of Borel
sets of diameters less than $1/n$ such that
$\mathfrak{S}(\mathcal{H} )=\bigcup_{i}A_{i}^{n}$, $A_{i}^{n}\cap
A_{j}^{n}=\emptyset $ for $j\neq i$. Find a number $\,m=m(n)$ such
that $\sum_{i=m+1}^{+\infty }\pi (A_{i}^{n})<1/n$. Consider the
finite collection of Borel set $\{\hat{A} _{i}^{n}\}_{i=1}^{m+1}$,
where $\hat{A}_{i}^{n}=A_{i}^{n}$ for all $i= \overline{1,m}$ and
$\hat{A}_{m+1}^{n}=\bigcup_{i=m+1}^{+\infty }A_{i}^{n}$. We have
\begin{equation}
\bar{\rho}(\pi )=\sum_{i=1}^{m+1}\int\limits_{\hat{A}_{i}^{n}}\rho
\pi (d\rho )=\sum_{i=1}^{m+1}\pi _{i}^{n}\rho _{i}^{n},
\label{int-decomp}
\end{equation}
where $\pi _{i}^{n}$
$=\mathrm{Tr}\int\limits_{\hat{A}_{i}^{n}}\rho \pi (d\rho )=\pi
(\hat{A}_{i}^{n})$ and $\rho _{i}^{n}=(\pi
(\hat{A}_{i}^{n}))^{-1}\int\limits_{\hat{A}_{i}^{n}}\rho \pi
(d\rho )$ (without loss of generality we assume $\pi _{i}^{n}>0$).
Let $\pi ^{n}$ be the probability measure on
$\mathfrak{S}(\mathcal{H})$, ascribing the value $\pi _{i}^{n}$ to
the point $\rho _{i}^{n}$. Equality (\ref{int-decomp}) implies
$\bar{\rho} (\pi ^{n})=\bar{\rho}(\pi )$. Since $\pi ^{n}$ has
finite support for each $ n $, to prove the assertion of the lemma
it is sufficient to show that $\pi ^{n}$ tends to $\pi $ in the
weak topology as $n$ tends to $+\infty $. By theorem 6.1 in
\cite{Par} to establish the above convergence it is sufficient to
show that
\[
\lim_{n\rightarrow +\infty }\int\limits_{\mathfrak{S}(\mathcal{H})}f(\rho
)\pi ^{n}(d\rho )=\int\limits_{\mathfrak{S}(\mathcal{H})}f(\rho )\pi (d\rho
)
\]
for arbitrary bounded uniformly continuous function $f(\rho )$ on $\mathfrak{
\ S}(\mathcal{H})$. Let $M_{f}=\sup_{\rho \in \mathfrak{S}(\mathcal{H}
)}|f(\rho )|$. For arbitrary $\,\varepsilon >0\;$ let $n_{\varepsilon }$ be
such that $\varepsilon n_{\varepsilon }>2M_{f}$ and
\[
\sup_{\rho \in U(n_{\varepsilon })}f(\rho )-\inf_{\rho \in U(n_{\varepsilon
})}f(\rho )<\varepsilon
\]
for arbitrary closed ball $U(n_{\varepsilon })$ of diameter $
1/n_{\varepsilon }.$Let $n\geq n_{\varepsilon }.$ By construction
the set $\hat{A}_{i}^{n}$ is contained in some ball $U_{i}(n)$ for
each $i=\overline{1,m}$. By (\ref{l1}) the state $\rho _{i}^{n}$
lies in the same ball $U_{i}(n)$. Hence we have
\[
\begin{array}{c}
|\int\limits_{\mathfrak{S}(\mathcal{H})}f(\rho )\pi ^{n}(d\rho
)-\int\limits_{\mathfrak{S}(\mathcal{H})}f(\rho )\pi (d\rho )| \\
\leq \sum\limits_{i=1}^{m+1}\int\limits_{\hat{A}_{i}^{n}}|f(\rho
)-f(\rho _{i})|\pi (d\rho ) \\ \leq \varepsilon \sum_{i=1}^{m}\pi
(\hat{A}_{i}^{n})+2M_{f}\pi
(\hat{A}_{m+1}^{n})<2\varepsilon .
\end{array}
\]
for all $n\geq n_{\varepsilon }$. $\square $

\textbf{B. Example of a channel without optimal generalized
ensembles.} Consider Abelian von Neumann algebra
$\mathbf{\textit{l}}_{\infty}$ and its predual
$\mathbf{\textit{l}}_{1}$. Let $\Phi$ be the noiseless channel on
$\mathbf{\textit{l}}_{1}$. Consider the sequence of states
$$\rho_{n}=
\{1-q_{n},\underbrace{\textstyle\frac{q_{n}}{n},\frac{q_{n}}{n},...,\frac{q_{n}}{n}}_{n},
0, 0 ...\},
$$
where $q_{n}$ is a sequence of numbers in $[0,1]$, which will be
defined below. Note that in this case
$\chi_{\Phi}(\rho_{n})=H(\rho_{n})=h_{2}(q_{n})+q_{n}\log n$,
where $h_{2}(x)=-x\log x-(1-x)\log(1-x)$. We will show later that
there exists the sequence $q_{n}$ such that
$\lim_{n\rightarrow+\infty}q_{n}=0$ while the corresponding
sequence $\chi_{\Phi}(\rho_{n})=H(\rho_{n})$ monotonously
increases to $1$. Let $q_{n}$ be such a sequence and $\mathcal{A}$
be the closure of the sequence $\rho_{n}$, which obviously
consists of states $\rho_{n}$ and pure state
$\rho_{*}=\lim_{n\rightarrow+\infty}\rho_{n}=\{1, 0, 0 ...\}$. By
definition and the above monotonicity $\bar{C}(\Phi
;\mathcal{A})=\lim_{n\rightarrow+\infty}\chi_{\Phi}(\rho_{n})=1$
while $\rho_{*}$ is the only optimal average state for the
$\mathcal{A}$-constrained channel $\Phi$ and
$\chi_{\Phi}(\rho_{*})=H(\rho_{*})=0$. So we have $\bar{C}(\Phi
;\mathcal{A})>\chi_{\Phi}(\rho_{*})$ and corollary 3 implies that
there is no optimal ensemble for the $\mathcal{A}$-constrained
channel $\Phi$.

Let us construct the sequence $q_{n}$ with the above properties.
Consider the strongly increasing function $f(x)=x(1-\ln x)$ on
$[0,1]$. It is easy to see that $f'(x)=-\ln x$ and
$f([0,1])=[0,1]$. Let $f^{-1}$ be the converse function and
$g(x)=xf^{-1}(\ln 2/x)$ for all $x\geq 1$. Note that the function
$g(x)$ is implicitly defined by the equation
\begin{equation}\label{a-e}
g(1-\ln(g/x))=\ln 2.
\end{equation}
Using this it is easy to see that the function $g(x)$ satisfies
the following differential equation
\begin{equation}\label{d-e}
\ln(g/x)g'=g/x.
\end{equation}
Since $g(x)/x=f^{-1}(\ln 2/x)$ we have $g(x)/x\in[0,1]$. This with
(\ref{a-e}) and (\ref{d-e}) implies $g(x)\in[0,1]$,
$\lim_{x\rightarrow+\infty}g(x)=0$ and $g'(x)<0$ correspondingly.
Consider the function $H(x)=h_{2}(g(x))+g(x)\log x$. By
(\ref{a-e}) and (\ref{d-e}) with the above observations we have
$$
\lim_{x\rightarrow+\infty}H(x)=(\ln
2)^{-1}\lim_{x\rightarrow+\infty}g(x)\ln x=1
$$
and
$$
\begin{array}{c}
H'(x)=(\ln 2)^{-1}\left(g'(x)\ln(1-g(x))-g'(x)\ln g(x)+g'(x)\ln
x+g(x)/x\right)\\\\=(\ln 2)^{-1}g'(x)\log (1-g(x))>0,\quad \forall
x>1.
\end{array}
$$
It follows that $H(x)$ is an increasing function on $[1,+\infty)$,
tending to its upper bound $1$ at infinity. Setting $q_{n}=g(n)$
we obtain the sequence with the desired properties.

{\bf Acknowledgements.}

The first author acknowledges support from QIS Program, the Newton Institute,
Cambridge, where this paper was completed. The work was partially supported by
INTAS grant 00-738.

\end{document}